\documentclass[a4paper,twoside]{article}

\usepackage{amsmath,amssymb,amsfonts}
\usepackage{algorithmic}
\usepackage{graphicx}
\usepackage{textcomp}
\usepackage[utf8]{inputenc}
\usepackage{multirow}
\usepackage[table,xcdraw]{xcolor}
\usepackage{todonotes}
\usepackage{rotating}
\usepackage{hyphenat}
\usepackage{varwidth}
\usepackage{makecell}
\usepackage{multirow}
\usepackage{colortbl}
\usepackage{hhline}
\usepackage{subfigure}
\usepackage{float}
\usepackage{caption}
\usepackage{booktabs}
\usepackage{makecell}
\usepackage{siunitx}
\usepackage{tcolorbox}
\usepackage{url}
\usepackage{xcolor}
\usepackage{hyperref}
\usepackage{float}

\usepackage{epsfig}
\usepackage{calc}
\usepackage{amstext}
\usepackage{amsthm}
\usepackage{multicol}
\usepackage{pslatex}
\usepackage{natbib}
\usepackage[bottom]{footmisc}
\usepackage{SCITEPRESS}     

\begin{document}

\title{CrudeBERT: Applying Economic Theory towards fine-tuning Transformer-based Sentiment Analysis Models to the Crude Oil Market}


\author{
    \authorname{Himmet Kaplan\sup{1}\orcidAuthor{0000-0002-1115-8669},  Ralf-Peter Mundani\sup{2}\orcidAuthor{0000-0001-6248-714X}, Heiko Rölke\sup{2}\orcidAuthor{0000-0002-9141-0886} and Albert Weichselbraun\sup{2}\orcidAuthor{0000-0001-6399-045X}}
    \affiliation{\sup{1}Zurich University of Applied Sciences, Winterthur, Switzerland}
    \affiliation{\sup{2}University of Applied Sciences of the Grisons, Chur, Switzerland}
}


\abstract{Predicting market movements based on the sentiment of news media has a long tradition in data analysis. With advances in natural language processing, transformer architectures have emerged that enable contextually aware sentiment classification. Nevertheless, current methods built for the general financial market such as FinBERT cannot distinguish asset-specific value-driving factors. This paper addresses this shortcoming by presenting a method that identifies and classifies events that impact supply and demand in the crude oil markets within a large corpus of relevant news headlines. We then introduce CrudeBERT, a new sentiment analysis model that draws upon these events to contextualize and fine-tune FinBERT, thereby yielding improved sentiment classifications for headlines related to the crude oil futures market. An extensive evaluation demonstrates that CrudeBERT outperforms proprietary and open-source solutions in the domain of crude oil.
}

\onecolumn \maketitle \normalsize \setcounter{footnote}{0} \vfill

\section{\uppercase{Introduction}}
\label{sec:introduction}
Crude oil is one of our primary energy sources and also one of the most influential raw materials. 
Thus, it is of utmost importance for the global economy and even serves as an indicator of economic boom or recession. 
Since crude oil is a limited natural resource, its price is expected to be determined by supply and demand. 
Yet, according to literature, crude oil is one of the most volatile markets in the world since its demand is primarily affected by economic activity (business cycle) and exogenous events such as armed conflicts and natural disasters \citep{buyuksahin_speculators_2011}.
Traditionally, analysts draw upon technical analysis which utilizes historical data for prediction. However, historical data rarely provides high-confidence insights \citep{mccarthy_applying_2019}. 
Complementing technical analysis with additional contemporary and relevant information such as news articles could be a promising strategy for achieving more reliable results. 
Several empirical studies demonstrated that considering news media significantly improved forecasts of large market movements (i.e., higher than 50\%) of publicly listed assets \citep{qian_stock_2007}. 
Therefore, many researchers studied the benefits of incorporating news data into their prediction models \citep{baboshkin_multi-source_2021}. 
One option for applying news to prediction tasks comes with sentiment analysis which quantifies the impact of news on a certain asset as positive, neutral, or negative. 
With the latest advancement in computer hard- and software, particularly the development of transformer architectures \citep{devlin_bert_2018}, modern natural language processing (NLP) algorithms emerged that are capable of evaluating text in a contextually aware manner for strategic forecasting. 
The observations of \citet{jiang_smart_2020} indicate that current transformer-based sentiment classifiers can achieve remarkable accuracies of up to 97.5\,\%. 
While these sentiment analysis methods gained great traction in prediction tasks for the stock market and cryptocurrencies, they still play only a minor role in forecasting crude oil prices. 
Thus, the benefits of considering the sentiment of news headlines for crude oil price predictions seem to be evident. 

The presented research draws upon FinBERT, a state-of-the-art transformer-based sentiment analysis model that has been pre-trained for the general financial market. 
However, analyzing over a decade of news headlines relevant to the crude oil market revealed that FinBERT's sentiment classification does not deliver any apparent insights into the contemporary development of oil prices.  
Therefore, we developed the publicly available CrudeBERT sentiment analysis model that has been optimized for the crude oil domain. CrudeBERT extends FinBERT by considering the economic theory of supply and demand. 
In our experiments, CrudeBERT outperforms FinBERT and provides a promising tool for improving crude oil price predictions by incorporating information on the sentiment conveyed in news headlines. 

The main contributions of this paper can be summarized as (i) developing a method that provides transformer models with means for identifying the major supply and demand factors that drive crude oil futures markets, (ii) fine-tuning general transformer-based sentiment analysis methods by incorporating the economic model of supply and demand into these models, and (iii) conducting extensive experiments that draw upon multiple prediction settings to benchmark the developed method against a baseline (random binary classification) and two state-of-the-art (lexicon- and transformer-based) sentiment analysis frameworks.

The remainder of this paper is organized as follows: Chapter~\ref{sec:related-work} discusses related literature that led to the modern transformer-era sentiment analysis applications. 
Afterward, chapter~\ref{sec:method} introduces the FinBERT domain-specific affective model for the domain of crude oil markets and its use in predicting market movements. 
Chapter~\ref{sec:evaluation} describes the evaluation setup, performs a comprehensive evaluation of the CrudeBERT model, and discusses the obtained results. Chapter~\ref{sec:outlook} concludes the paper with a summary and an outlook on future improvements.

\section{\uppercase{Related Work}}
\label{sec:related-work}
From an industrial standpoint, crude oil is critical to the world's economy. Consequently, many research articles focus on predicting its price with studies varying from technical to fundamental analysis. This literature review focuses on articles aimed at forecasting crude oil prices by including sentiment features.

\subsection{Efficient Market Hypothesis}
The efficient market hypothesis (EMH) questions whether information retrieved from news articles does contain any predictive value at all since it claims that the price of an asset already considers all publicly available information. 
Eugene Fama distinguishes between the weak, semi-strong, and strong forms of EMH \citep{fama_efficient_1970}.  
The weak form claims that the price results only from its historical price history, thus making all available information outside the historical price relevant for forecasting the future price of an asset. The EMH's semi-strong variant on the other hand considers that the current pricing reflects the historical prices and publicly available information. Therefore, confidential information such as insider knowledge can add value to a prediction, given it hasn't yet altered the current price \citep{malkiel_efficient_1989}.
The strong form of EMH assumes that prices reflect historical prices, and publicly available, and confidential information \citep{fama_efficient_1970}. Hence, the strong form assumes that applying fundamental analysis based on any available information cannot lead to abnormal economic returns. This form of the EMH is further supported by various studies that emphasize the notorious difficulty of forecasting crude oil prices, such as the works of Hamilton, which concludes that the oil price appears to be influenced by a random walk with drift \citep{hamilton_understanding_2008}. 

Yet, numerous experts have questioned the hypothesis of the EMH's strong and semi-strong forms, claiming that once a news message is published, the available information changes, and, therefore, the price is expected to adapt. In the experiments of Qian and Rasheed, they concluded that the predictions based on news can correctly forecast price fluctuations with greater than 50\,\% accuracy \citep{qian_stock_2007}.
Furthermore, according to Buyuksahin and Harris, crude oil demand is primarily driven by exogenous events, such as armed conflicts and natural disasters as well as the presence of speculators such as noise traders. They assume that these events considerably contribute towards making crude oil one of the most volatile markets in the world. They also observed a substantial relationship between crude oil price changes and the behavior of politically and economically unstable nations, which often trigger such exogenous events \citep{buyuksahin_speculators_2011}. This observation is confirmed by Brandt and Gao's more recent study, which shows that macro\-economic and geopolitical news has a strong influence on crude oil, with varying impacts. For example, macroeconomic news influences short-term price movements and also helps to forecast long-term oil prices. On the other hand, the influence of geopolitical news yields typically a robust and instantaneous impact that results in increased volume in trade. However, geopolitical news delivers no conclusive insights in terms of forecasting \citep{brandt_macro_2019}. Wex et al. state that forecasts based on the sentiment scores of news articles that cover exogenous events are statistically significant \citep{wex_early_2013}.

\subsection{Sentiment Analysis}
Sentiment analysis is considered a prevalent classification task in NLP, which categorizes affective and subjective information within entire documents, paragraphs, and sentences. 
It has gained increasing popularity due to its vast potential for a variety of applications such as economics, finance, marketing, political science, psychology, and human-computer interaction \citep{mohammad_sentiment_2021}. 
Sentiment in the context of sentiment analysis, which is also known as opinion mining, refers to the quantification of natural language within pre-defined affective dimensions \citep{weichselbraun_automatic_2022} such as sentiment polarity which distinguishes between positive, neutral, and negative media coverage. Still, there is little discussion about what sentiment in the context of NLP truly represents \citep{hovy_what_2015}. Generally, researchers assume that authors always express some sentiment while producing natural language, since emotions, opinions, and expressions in language are fundamental human traits \citep{Taboada2016}. Therefore, sentiment analysis can also cover complex emotions such as the ones introduced in Plutchik's Wheel of Emotions \citep{plutchik_psychoevolutionary_1982} and the Hourglass of Emotions \citep{susanto_hourglass_2020}. However, most literature in finance tends to break sentiment down into attitudes using binary polarities such as positive and negative, also known as financial sentiment analysis (FSA). Hence a binary classification is more suitable for directly assessing the up or down movements of publicly traded assets \citep{li_news_2014}.

\subsection{Early NLP Methods for FSA}
Sentiment analysis in the financial domain has been introduced in the 1980s. One of the first approaches to classifying sentiment in text documents was the Bag-of-Words (BOW) methodology, often referred to as the lexicon-based technique \citep{liew_fine-grained_nodate}. Since a text consists of several words (tokens), BOW simply accumulates the sentiment scores of positive and negative words to compute the overall sentiment classification. As the name suggests, these BOW methods utilize a lexicon consisting of words and their sentimental value, predetermined by, ideally multiple, human annotators. One of the most well-known lexicons for FSA was developed by Loughran and McDonald and aimed at interpreting liabilities concerning 10-K filing returns \citep{loughran_when_2011}. In their later works \citep{loughran2016textual}, they published a survey on the use of text analysis with a focus on accounting and finance. However, creating lexicons that include all possible keywords including negates, or word combinations is very challenging, since a term's sentiment often also depends on the context expressed by surrounding terms or paragraphs.

\subsection{Machine Learning-based Sentiment Analysis}
With advancements in computer hard- and software, modern FSA approaches started to heavily rely on machine learning-based approaches. These approaches mostly focused on supervised learning, in which learning is accomplished by training on annotated datasets containing pairs of inputs and matching solutions. 
As a result, by correcting and optimizing themselves based on the, mostly human-curated, training dataset the machine learning-based approaches identify rules and patterns and attempt to derive a potentially meaningful generalization. These approaches, which require large amounts of annotated training data, are known as supervised learning and are usually used for classification and regression \citep{chollet_deep_2018} tasks. For instance, Recurrent Neural Networks (RNN) and Long Short-Term Memory (LSTM) are particularly well suited for sequential data, such as text \citep{tang-etal-2016-effective}. 
However, the required training dataset is one disadvantage of supervised learning, particularly for classification problems such as sentiment analysis. Since a bigger training dataset usually yields better results using large training datasets is typically expensive (both computationally and financially).
Furthermore, RNNs suffer from vanishing and exploding gradients, making them unsuitable for lengthy texts, and are slower to train since their sequential flow is incompatible with parallel processing \citep{chollet_deep_2018}. 
One approach to addressing this problem is combining supervised approaches with unsupervised machine learning models.
For instance, word embeddings (also known as word vector models), map words into a vector space that aligns semantically related words close to each other in an unsupervised manner. 
Among the most popular word vector models are word2vec \citep{mikolov_2013} and Glove \citep{pennington_glove_2014} which can be trained on large text corpora, therefore, capturing the word semantics within the corpus. Nevertheless, word embeddings still lack the power to fully consider a term's context -- i.e., once a model has been trained, words with the same spelling always receive the same word vector, independent of their context -- i.e., orange either as a fruit or as a color, or a mixture of both concepts.

\subsection{Transformer-based Sentiment Analysis}
\begin{figure}[htb]
    \includegraphics[width=\columnwidth]{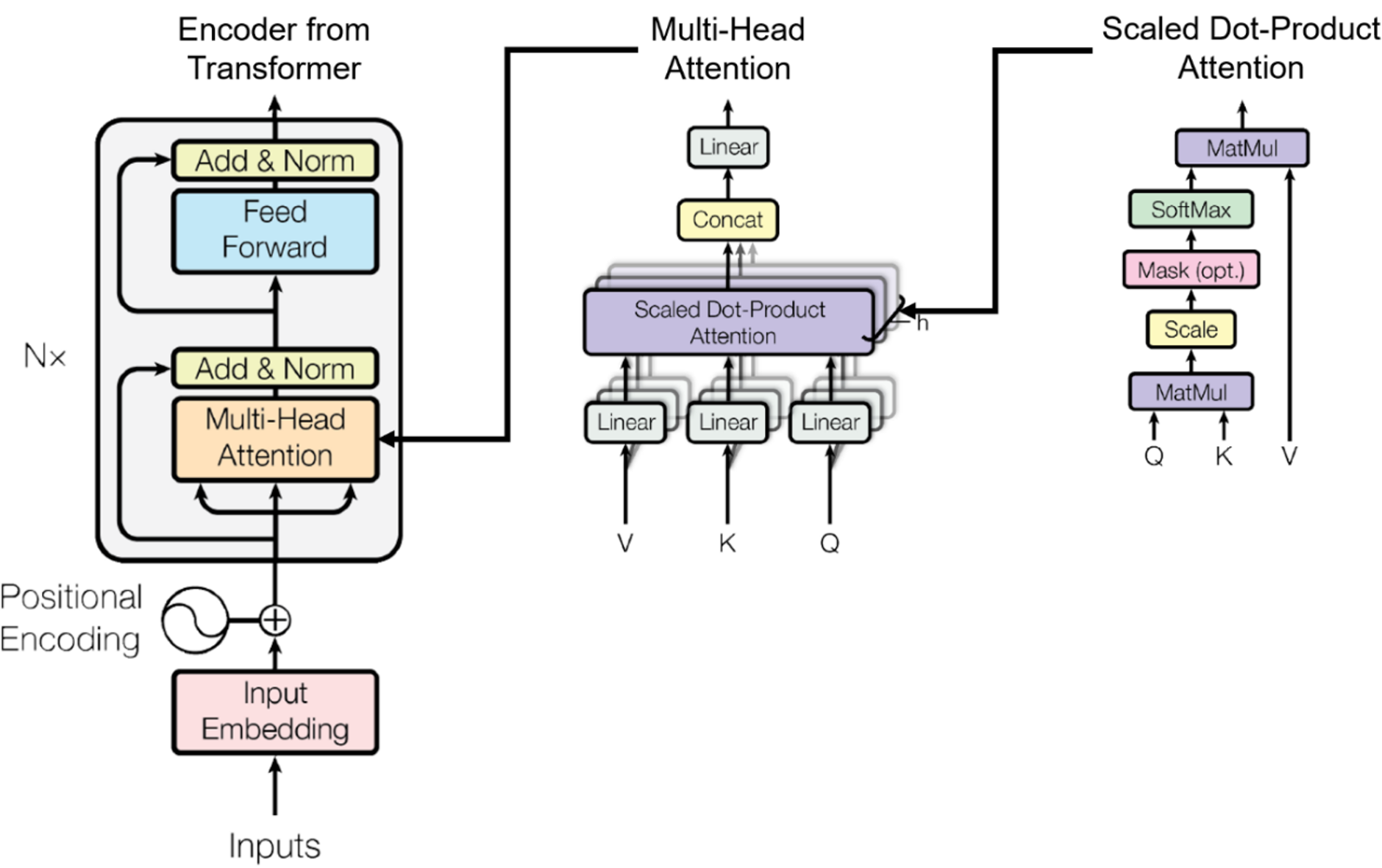}
    \caption{\label{fig:components-attention}Components of the Multi-Head Attention Design. \citep{vaswani_attention_2017}}
\end{figure}

More recent language models draw upon the attention mechanism \citep{bahdanau_neural_2016} which led to significant gains in a range of NLP tasks including sentiment analysis. The attention mechanism allows neural networks to resemble the human cognitive function by selectively focusing on particularly relevant information while dismissing other less relevant information. This approach encourages the neural network to spend more computational resources on small but relevant elements of the data \citep{bahdanau_neural_2016} yielding improvements in terms of speed and accuracy. 
Further enhancements from Vaswani et al. made use of the attention mechanism to develop the transformer architecture, which allows parallel training and makes it more efficient than RNNs. Initially, the transformer architecture was proposed for neural machine translation, thus, it contains an encoder and decoder. The encoder is a fully connected feed-forward network made out of multiple identical multi-headed attention layers which allows the sequence to be evaluated from contextually varying perspectives (Figure~\ref{fig:components-attention}).
The capability to consider a term's context together with the option to draw upon and customize large pre-trained models has been key to the success of transformer-based language models for sentiment analysis \citep{vaswani_attention_2017}.

\subsection{FinBERT for Financial Sentiment Analysis}
\begin{figure}[h]
    \includegraphics[width=\columnwidth]{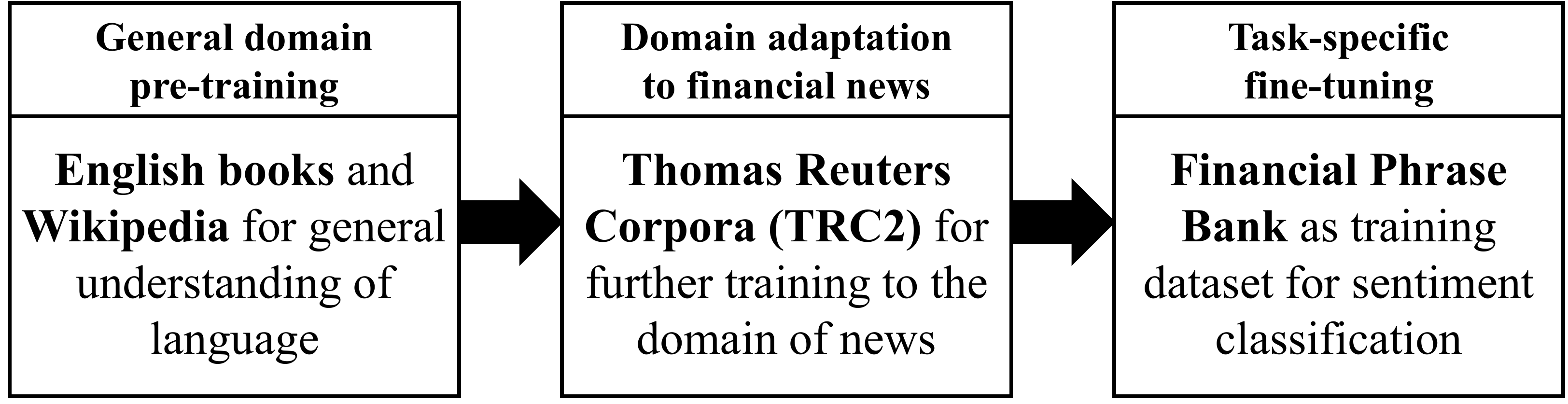}
    \caption{\label{fig:finbert-process}Process of generating FinBERT.}
\end{figure}

Shortly after the release of the transformer architecture Develin et al. observed that the encoder, when layered, can also serve as a strong representation learning model and for this matter, they developed the Bidirectional Encoder Representations from Transformers (BERT) \citep{devlin_bert_2018}. One noteworthy feature of BERT was its simple customization for a wide range of NLP tasks with the important added capability of contextual perception of words \citep{yenicelik_understanding_2020}. Initially, it was pre-trained on Eng\-lish \textit{Wikipedia} and the \textit{BookCorpus} to give the model a general comprehension of natural language \citep{zhu_aligning_2015}. This model served as a foundation for further adaptions to specific NLP applications and its domain, such as FinBERT \citep{araci_finbert_2019} which focuses on sentiment analysis of financial news. To achieve this, Araci et al. employed a subset of the \textit{Thomson Reuters} Text Research Collection (TRC2) to adapt the model to the domain of financial news, where occurrences of slang and spelling errors are minimal. For the task-specific fine-tuning process, the training dataset \textit{Financial Phrase Bank} from \citet{malo_good_2014} was utilized (Figure~\ref{fig:finbert-process}).

Compared to the number of papers that used FinBERT as a backend for classification, the proportion of papers that use it for classifying sentiments towards crude oil is relatively small.

\subsection{RavenPack Event Sentiment Score}
To capture the overall sentiment of the market, \textit{RavenPack} developed a lexicon-based news sentiment index namely the \textit{Event Sentiment Score} (ESS), which is a granular score between -1 (negative sentiment) and 1 (positive sentiment).
This score is determined by systematically comparing stories that are often classified as having a good or negative financial or economic impact through manual assessments by experts. Based on this human-curated lexicon the ESS algorithm can examine a wide range of sentiment proxies that are frequently mentioned in financial news allowing it to classify the sentiment from earnings reports to natural disasters. \citep{hafez_ess_factor_2020}

\section{\uppercase{Method}}
\label{sec:method}
This section introduces the CrudeBERT model, which extends FinBERT by incorporating knowledge of an event's expected impact on crude oil supply and demand.
Section~\ref{sec:method-datasets} presents an overview of the used news headlines and crude oil price data sets which is followed by a discussion of the data pre-processing steps. Afterward, we analyse the shortcomings and flaws of FinBERT and addressed them by developing the CrudeBERT model.

\subsection{Datasets}
\label{sec:method-datasets}

\subsubsection{News Data}
\label{sec:method-news-data}
The dataset containing news information consists of around 46,000 headlines published between 1 January 2000, and 1 April 2021, with high relevance to the topic of crude oil and obtained through the \textit{RavenPack Realtime news Discovery} platform. Similar to Li et al., we limit our analysis to news headlines, since they are more easily accessible, and have lower requirements in terms of pre-processing, storage, and computational power \citep{li_text-based_2019}. The headlines used in this paper originate from 1034 unique news sources, of which the majority has been published on the \textit{Dow Jones newswires} (approx. 21,200), followed by \textit{Reuters} (approx. 3,000), \textit{Bloomberg} (approx. 1,100), and \textit{Platts} (approx. 870). There are also around 400 news sources present that only delivered a single headline. It should be noted that RavenPack has added new publishers over the years, which led to a steady increase in the number of available sources over time, especially till 2012. To ensure rich news coverage with diverse sources, we limit our evaluations to the period after 2012.

\subsubsection{Price Data}
\label{sec:method-price-data}
The oil market is dominated by the two most prevalent grades Brent Crude and Western Texas Intermediate (WTI), which dictate the price of crude oil \citep{us_energy_information_administration_table_nodate}. Brent crude is the benchmark for crude oil in Africa, Europe, and the Middle East, accounting for almost two-thirds of the global supply. WTI, on the other hand, is the favored benchmark used by the United States of America. Since all of the headlines in our dataset are in English, the WTI futures prices were regarded as potentially more relevant for our research. The historical values of WTI have been acquired from the financial market platform \textit{investing.com} for the same period as the headlines.

\subsection{Data Pre-processing}
\subsubsection{Sentiment Data Normalization}
We normalized the sentiment values of headlines, computed by the sentiment classifiers, by using z-statistics with the aim to integrate the market's relative mood into the classification model. By normalizing sentiment data over a sliding window we account for the perfect market theory (i.e., the market price reflects all publicly available information) by assuming that only new information that either disappoints or excels stakeholder expectations results in significant price changes.

Equation~\ref{eq:sent_norm} outlines the normalization of the sentiment value at time point $t$ based on a weekly sliding window of size $w =5$ with
\begin{eqnarray}
\label{eq:sent_norm}
  sent_{norm, t} &=& \frac{sent_{t}-\overline{sent}_{t,w}}{\sigma_{t,w}} 
\end{eqnarray}
where $\overline{sent}_{t, w}$ indicates the average sentiment at time points $t, t-1, ... t-w$ within the sliding window, and $\sigma_{t,w}$ the corresponding standard deviation.

\subsubsection{Price Data Normalization}
Due to market volatility, commodity and stock prices show random fluctuations that overlap short-term and long-term trends within the market. Therefore, we also normalize price data using z-statistics to better distinguish between significant market movements and random fluctuations. As with the $sent_{norm, t}$ we normalize price data for a weekly sliding window of $w = 5$ (due to the market being closed over the weekends) as outlined in the equation below:

\begin{eqnarray}
  price_{norm,t} &=& \frac{price_t-\overline{price}_{t,w}}{\sigma_{t,w}} 
\end{eqnarray}
with $\overline{price}_{t,w}$ and $\sigma_{t,w}$ indicating the average price and standard deviations within the chosen sliding window.

\subsubsection{Handling Multiple Daily Sentiment Scores}
Days covered by multiple news headlines yield multiple sentiment scores, which need to be merged for that given day. 
Prior work by Hafez et al. concluded that using the sum rather than the mean can examine both the sentiment score and the sentiment volume at the same time, even though, the normalization of scores would severely shrink the relative impact of days with a lower news volume.
According to their research, this strategy resulted in superior outcomes in their experiments \citep{hafez_effects_2018}. 

\subsubsection{Handling Gaps in the Dataset}
Rows containing gaps caused either by missing headlines or missing prices (due to closings of the market) were dropped entirely as a row. Furthermore, all the values have been scaled between $-1$ and $1$. The final dataset covers the period from 1 January 2012 to 1 April 2021 and yields 3376 rows of data. 

The dataset aligns the summarized daily sentiment scores with the price change of the following day ($Return_{t+1}$), i.e., assumes that markets will adapt to new information by the next day at the latest.
Sentiment scores vary between positive ($1$) and negative ($-1$) values. The price, in contrast, always remains positive with the notable exception of 20 April 2020 when prices became negative for a short period. 
Price changes (i.e., $Returns$) are, therefore, better suited for indicating the market's reaction to news coverage.
We compute the daily $Returns$ of WTI crude oil futures as outlined in Equation~\ref{eq:return} and compare them to the sentiment scores.

\begin{eqnarray}
	\label{eq:return}
  Return &=& \frac{Price_t-{Price}_{t-1}}{{Price}_{t-1}} 
\end{eqnarray}

Interpreting the oil price as the result of cumulative returns allows a comparison to the cumulative sentiment scores, as illustrated in Figure~\ref{fig:plot_cumulative_values}.

\subsection{Shortcomings of the FinBERT}
\label{sec:method-finbert}
Figure~\ref{fig:plot_cumulative_values} performs a visual comparison of FinBERT's cumulative sentiment scores (red) and the price (blue) history of WTI crude oil futures to assess FinBERT's forecasting potential. The plot does not show any apparent relationship or trends and outlines the need for an additional inquiry into the underlying causes of this poor relationship and potential ways for correcting it.

Adam Smith's (1776) price theory advocates that the price of limited resources such as crude oil is determined by supply and demand. 
In this context, \textit{supply} refers to the amount of a product or service that a provider will sell at a given price during a specific period. 
\textit{Demand} denotes the amount of a product or service that a buyer is willing to acquire during the same period for a given price. 
The interaction between suppliers and customers yields a competitive market in which the price of products and services is determined by the equilibrium between supply and demand \citep{smith_inquiry_1776}.
For example, if demand remains constant but supply falls, the resulting shortage will cause prices to rise. A shortage can  also occur if the supply remains constant but the demand rises. 

In contrast, increased supply with constant demand will result in a surplus and consequently a decrease in prices. A surplus can also emerge if supply remains constant but demand falls. This logic behind supply and demand can be summed up as follows: 

Less supply $\rightarrow$ shortage $\rightarrow$ higher price 

More supply $\rightarrow$ surplus $\rightarrow$ lower price

Less demand $\rightarrow$ surplus $\rightarrow$ lower price

More demand $\rightarrow$ shortage $\rightarrow$ higher price

A drill-down analysis that compared news headlines to FinBERT sentiment scores revealed that FinBERT tended to produce dubious outcomes. Given that crude oil is a publicly-traded asset and FinBERT has been trained on general financial market news this result seems arguably surprising. Having said that, according to Xing et al. such behavior is expected when utilizing general sentiment analysis methods for a specific domain and is known as the domain adaptation problem \citep{xing_financial_2020}. \citet{weichselbraun_automatic_2022} also emphasize the need for domain-specific affective models and present methods for creating such models.

Interpreting the FinBERT scores of news headlines listed in Table~\ref{tab:should-finbert} based on the impact of supply and demand on prices reveals some serious issues with FinBERT's assessment of strongly positive (+1), highly negative (-1), and neutral (0) events.
Headlines suggesting a drop in supply (e.g., due to accidents at oil refineries and oil platforms), tend to ensue negative FinBERT scores although the corresponding events likely lead to higher crude oil prices. The FinBERT model probably returns these negative scores since accidents are rarely good news in finance and due to moral assessments derived from the human-made annotations within the \textit{Financial Phrase Bank}.

Headlines implying a rise in supply (e.g., due to oil discoveries and increasing exports), in contrast, frequently yield neutral FinBERT scores.

When it comes to a decline in demand (e.g., induced by decreased imports), the resulting surplus should lead to a price decrease. This assessment is also confirmed by two of the three FinBERT scores for the analyzed headlines (row supply \emph{surplus} due to \emph{decreasing demand}) in Table~\ref{tab:should-finbert}. 
The first headline, in contrast, yields a positive FinBERT score, since FinBERT is not able to correctly interpret the fall in imports indicated by negative numbers such as -16.0\,\%. This limitation should be taken under consideration when utilizing FinBERT since a substantial number of headlines do contain such values. 
Lastly, headlines that indicate increasing demand should result in higher oil prices. The experiments with FinBERT confirm that it considers headlines conveying increased demand mostly as positive.

\begin{table}[htb]
\caption{\label{tab:should-finbert}Sample of headlines and output of FinBERT.}
\centering
\resizebox{\columnwidth}{!}{%
\begin{tabular}{|c|c|c|c|c|} 
\hline
\multicolumn{3}{|c|}{\textbf{Headlines}}                                                                                                                                                                                         & \begin{tabular}[c]{@{}c@{}}\textbf{Sentiment }\\\textbf{Score }\\\textbf{Expected}\end{tabular} & \begin{tabular}[c]{@{}c@{}}\textbf{Sentiment }\\\textbf{Score }\\\textbf{FinBERT}\end{tabular}  \\ 
\hline
\multirow{14}{*}{\rotcell{\nohyphens{\textbf{Shortage}}}} & \multirow{7}{*}{\rotcell{\nohyphens{\textbf{Supply Decrease}}}} & \begin{tabular}[c]{@{}c@{}}\textbf{Major Explosion, Fire at Oil }\\\textbf{Refinery in Southeast Philadelphia}\end{tabular} & {\cellcolor[rgb]{0.388,0.745,0.482}}Positive                                                    & {\cellcolor[rgb]{0.973,0.447,0.424}}-0.886292                                                   \\ 
\hhline{|~~---|}
                                             &                                                     & \begin{tabular}[c]{@{}c@{}}\textbf{PETROLEOS confirms Gulf of }\\\textbf{Mexico oil platform accident}\end{tabular}         & {\cellcolor[rgb]{0.388,0.745,0.482}}Positive                                                    & {\cellcolor[rgb]{0.984,0.631,0.459}}-0.507213                                                   \\ 
\hhline{|~~---|}
                                             &                                                     & \begin{tabular}[c]{@{}c@{}}\textbf{CASUALTIES FEARED AT OIL }\\\textbf{ACCIDENT NEAR IRANS BORDER }\end{tabular}            & {\cellcolor[rgb]{0.388,0.745,0.482}}Positive                                                    & {\cellcolor[rgb]{0.973,0.439,0.424}}-0.901763                                                   \\ 
\hhline{|~----|}
                                             & \multirow{7}{*}{\rotcell{\nohyphens{\textbf{Demand Increase}}}} & \begin{tabular}[c]{@{}c@{}}\textbf{EIA Chief expects Global Oil }\\\textbf{Demand Growth 1 M B/D to 2011}\end{tabular}      & {\cellcolor[rgb]{0.388,0.745,0.482}}Positive                                                    & {\cellcolor[rgb]{0.388,0.745,0.482}}0.930822                                                    \\ 
\hhline{|~~---|}
                                             &                                                     & \begin{tabular}[c]{@{}c@{}}\textbf{Turkey Jan-Oct Crude }\\\textbf{Imports +98.5\% To 57.9M MT}\end{tabular}                & {\cellcolor[rgb]{0.388,0.745,0.482}}Positive                                                    & {\cellcolor[rgb]{0.435,0.761,0.486}}0.866315                                                    \\ 
\hhline{|~~---|}
                                             &                                                     & \begin{tabular}[c]{@{}c@{}}\textbf{China's crude oil imports }\\\textbf{up 78.30\% in February 2019}\end{tabular}           & {\cellcolor[rgb]{0.388,0.745,0.482}}Positive                                                    & {\cellcolor[rgb]{0.396,0.749,0.486}}0.922963                                                    \\ 
\hline
\multirow{14}{*}{\rotcell{\nohyphens{\textbf{Surplus}}}}  & \multirow{7}{*}{\rotcell{\nohyphens{\textbf{Demand Decrease}}}} & \begin{tabular}[c]{@{}c@{}}\textbf{China February Crude }\\\textbf{Imports -16.0\% On Year }\end{tabular}                   & {\cellcolor[rgb]{0.973,0.412,0.42}}Negative                                                     & {\cellcolor[rgb]{0.671,0.827,0.502}}0.540711                                                    \\ 
\hhline{|~~---|}
                                             &                                                     & \begin{tabular}[c]{@{}c@{}}\textbf{Turkey May Crude Imports }\\\textbf{down 11.0\% On Year}\end{tabular}                    & {\cellcolor[rgb]{0.973,0.412,0.42}}Negative                                                     & {\cellcolor[rgb]{0.973,0.412,0.42}}-0.965965                                                    \\ 
\hhline{|~~---|}
                                             &                                                     & \begin{tabular}[c]{@{}c@{}}\textbf{Japan June Crude Oil Imports }\\\textbf{decrease 10.9\% On Yr}\end{tabular}              & {\cellcolor[rgb]{0.973,0.412,0.42}}Negative                                                     & {\cellcolor[rgb]{0.973,0.416,0.42}}-0.955271                                                    \\ 
\hhline{|~----|}
                                             & \multirow{6}{*}{\rotcell{\nohyphens{\textbf{Supply Increase}}}} & \begin{tabular}[c]{@{}c@{}}\textbf{Iran’s' Feb Oil Exports +20.9\% }\\\textbf{On Mo at 1.56M B/D - Official}\end{tabular}   & {\cellcolor[rgb]{0.973,0.412,0.42}}Negative                                                     & {\cellcolor[rgb]{0.961,0.914,0.518}}0.139093                                                    \\ 
\hhline{|~~---|}
                                             &                                                     & \begin{tabular}[c]{@{}c@{}}\textbf{Apache announces large petroleum }\\\textbf{discovery in Philadelphia}\end{tabular}      & {\cellcolor[rgb]{0.973,0.412,0.42}}Negative                                                     & {\cellcolor[rgb]{0.996,0.922,0.518}}0.089624                                                    \\ 
\hhline{|~~---|}
                                             &                                                     & \begin{tabular}[c]{@{}c@{}}\textbf{Turkey finds oil near }\\\textbf{Syria, Iraq border}\end{tabular}                        & {\cellcolor[rgb]{0.973,0.412,0.42}}Negative                                                     & {\cellcolor[rgb]{0.996,0.918,0.514}}0.076210                                                    \\
\hline

\end{tabular}%
}
\end{table}

\subsection{Extending FinBERT to CrudeBERT}
In the next step, we extended FinBERT to CrudeBERT to consider the economic law of supply and demand in the model's assessment.

\subsubsection{Training dataset generation}
To generate a domain-specific labeled silver standard for CrudeBERT, we analyzed several hundred headlines to determine frequently recurring topics, keywords indicating these topics, and their likely impact on the supply and demand of crude oil. 
This process identified the following major topics: \textit{accidents, oil discoveries, changes in exports, changes in imports, changes in demand, pricing, supply, pipeline limitations, drilling,} and \textit{spillage}. 

\begin{figure}[htb]
\begin{varwidth}{\linewidth}
    \resizebox{\columnwidth}{!}{\includegraphics[width=0.5\textwidth]{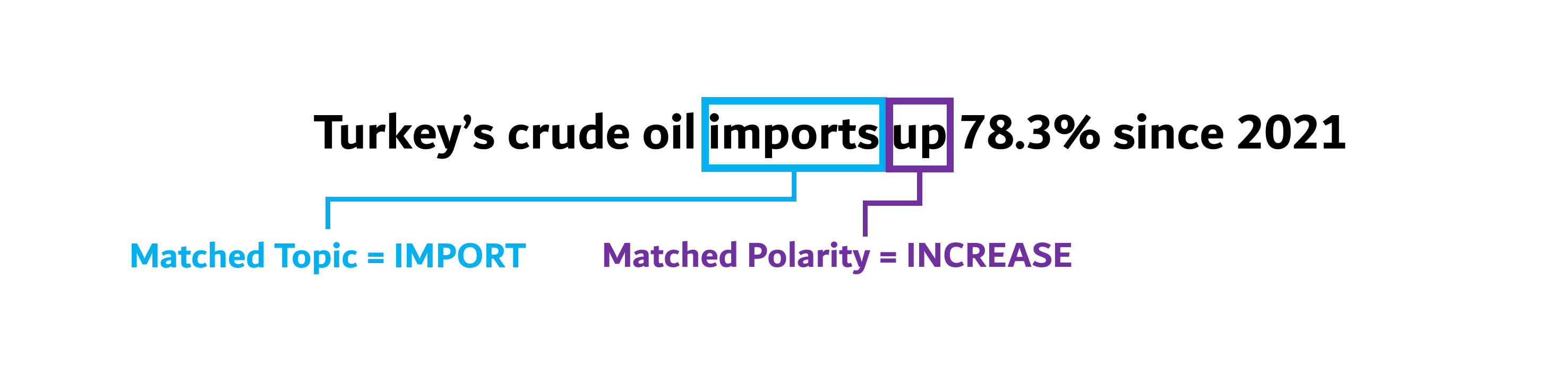}}
    \caption{\label{fig:Identifying Topics}Example of Detected Topic and Polarity.}
\end{varwidth}    
\end{figure}

We then queried the RavenPack repository for headlines containing the identified keywords and assigned them to the corresponding topic.
The headline in Figure~\ref{fig:Identifying Topics}, for instance, was assigned the topic \emph{changes in imports} due to the occurrence of the word \textit{import} in the headline. 
Afterward, we determined the direction of the change by classifying the headline's polarity based on the presence of terms that indicate an increase, a decrease, or constant levels.

The described approach enabled us to provide topic and direction labels for around 30,000 headlines. 
Table~\ref{tab:reoccurring-topics} summarizes the ten detected frequently reoccurring topics and their corresponding frequencies (we do not report the number of headlines with overlapping topics since it has been negligibly small):

\begin{table}[htb]
	\caption{\label{tab:reoccurring-topics}Frequency of Reoccurring Topics in the Domain of Crude Oil.}
\resizebox{\columnwidth}{!}{%
\begin{tabular}{|
>{\columncolor[HTML]{FFFFFF}}c 
>{\columncolor[HTML]{FFFFFF}}c 
>{\columncolor[HTML]{FFFFFF}}c 
>{\columncolor[HTML]{FFFFFF}}c |
>{\columncolor[HTML]{FFFFFF}}c 
>{\columncolor[HTML]{FFFFFF}}c 
>{\columncolor[HTML]{FFFFFF}}c 
>{\columncolor[HTML]{FFFFFF}}c |}
\hline
\multicolumn{4}{|c|}{\cellcolor[HTML]{FFFFFF}\textbf{Supply change}} &
  \multicolumn{4}{c|}{\cellcolor[HTML]{FFFFFF}\textbf{Demand change}} \\ \hline
\multicolumn{1}{|c|}{\cellcolor[HTML]{FFFFFF}Increase} &
  \multicolumn{2}{c|}{\cellcolor[HTML]{FFFFFF}No change} &
  Decrease &
  \multicolumn{1}{c|}{\cellcolor[HTML]{FFFFFF}Increase} &
  \multicolumn{2}{c|}{\cellcolor[HTML]{FFFFFF}No change} &
  Decrease \\ \hline
\multicolumn{1}{|c|}{\cellcolor[HTML]{FFFFFF}ca. 5900} &
  \multicolumn{2}{c|}{\cellcolor[HTML]{FFFFFF}ca. 350} &
  ca. 5700 &
  \multicolumn{1}{c|}{\cellcolor[HTML]{FFFFFF}ca. 1300} &
  \multicolumn{2}{c|}{\cellcolor[HTML]{FFFFFF}ca. 50} &
  ca. 800 \\ \hline
\multicolumn{4}{|c|}{\cellcolor[HTML]{FFFFFF}\textbf{Export change}} &
  \multicolumn{4}{c|}{\cellcolor[HTML]{FFFFFF}\textbf{Import change}} \\ \hline
\multicolumn{1}{|c|}{\cellcolor[HTML]{FFFFFF}Increase} &
  \multicolumn{2}{c|}{\cellcolor[HTML]{FFFFFF}No change} &
  Decrease &
  \multicolumn{1}{c|}{\cellcolor[HTML]{FFFFFF}Increase} &
  \multicolumn{2}{c|}{\cellcolor[HTML]{FFFFFF}No change} &
  Decrease \\ \hline
\multicolumn{1}{|c|}{\cellcolor[HTML]{FFFFFF}ca. 2000} &
  \multicolumn{2}{c|}{\cellcolor[HTML]{FFFFFF}ca. 150} &
  ca. 1500 &
  \multicolumn{1}{c|}{\cellcolor[HTML]{FFFFFF}ca. 2800} &
  \multicolumn{2}{c|}{\cellcolor[HTML]{FFFFFF}ca. 50} &
  ca. 2300 \\ \hline
\multicolumn{4}{|c|}{\cellcolor[HTML]{FFFFFF}\textbf{Price change}} &
  \multicolumn{1}{c|}{\cellcolor[HTML]{FFFFFF}\textbf{Spill}} &
  \multicolumn{2}{c|}{\cellcolor[HTML]{FFFFFF}\textbf{Discovery}} &
  \textbf{Drilling} \\ \hline
\multicolumn{2}{|c|}{\cellcolor[HTML]{FFFFFF}Increase} &
  \multicolumn{2}{c|}{\cellcolor[HTML]{FFFFFF}Decrease} &
  \multicolumn{1}{c|}{\cellcolor[HTML]{FFFFFF}ca. 2300} &
  \multicolumn{2}{c|}{\cellcolor[HTML]{FFFFFF}ca. 1600} &
  ca. 100 \\ \hline
\multicolumn{2}{|c|}{\cellcolor[HTML]{FFFFFF}} &
  \multicolumn{2}{c|}{\cellcolor[HTML]{FFFFFF}} &
  \multicolumn{2}{c|}{\cellcolor[HTML]{FFFFFF}\textbf{Accident}} &
  \multicolumn{2}{c|}{\cellcolor[HTML]{FFFFFF}\textbf{Pipeline issue}} \\ \cline{5-8} 
\multicolumn{2}{|c|}{\multirow{-2}{*}{\cellcolor[HTML]{FFFFFF}ca. 1600}} &
  \multicolumn{2}{c|}{\multirow{-2}{*}{\cellcolor[HTML]{FFFFFF}ca. 1300}} &
  \multicolumn{2}{c|}{\cellcolor[HTML]{FFFFFF}ca. 400} &
  \multicolumn{2}{c|}{\cellcolor[HTML]{FFFFFF}ca. 100} \\ \hline
\end{tabular}
}
\end{table}

\begin{figure}[H]
\begin{varwidth}{\linewidth}
    \resizebox{\columnwidth}{!}{\includegraphics[width=0.5\textwidth]{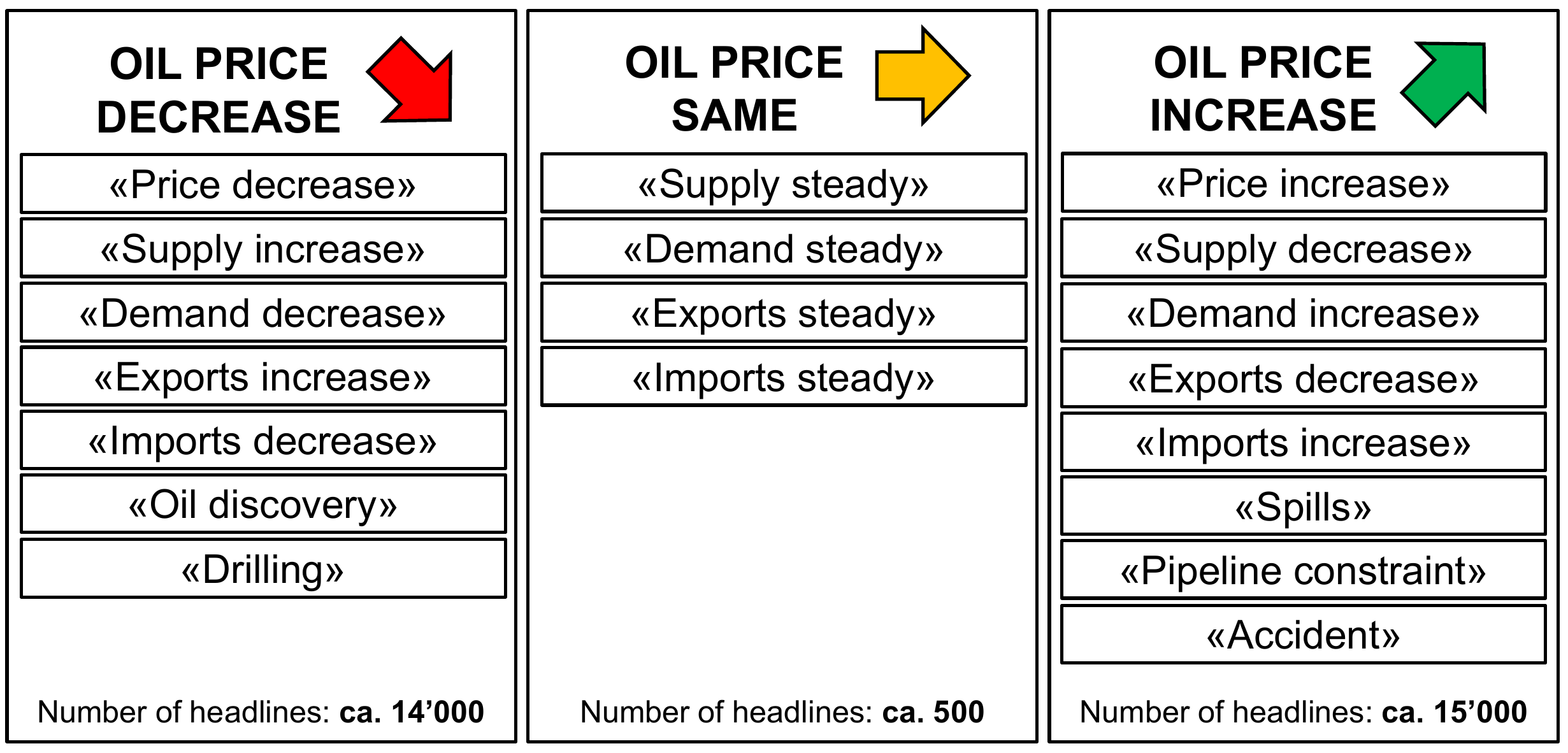}}
    \caption{\label{fig:sd-dataset}Assignment of the Labelled Topics.}
\end{varwidth}    
\end{figure}

Assessing these labels based on the price theory of supply and demand, allowed the creation of a domain-specific silver standard that classifies the headlines into positive (i.e., indicating increasing crude oil prices), negative (i.e., likely to cause decreasing crude oil prices), and neutral (i.e., should not affect the crude oil price), as outlined below (Figure~\ref{fig:sd-dataset}):

\begin{itemize}
	\item \emph{Lower prices (score: $-1$):} Headlines covering events such as drilling, discovery, increased exports, or simply a rise in oil production are likely to cause an increase in supply. 
		Similarly, headlines stating that oil imports or consumptions are decreasing should, in principle, result in a surplus of oil and, therefore, lower prices.
\item \emph{Higher prices (score: $+1$):} Headlines announcing accidents, pipeline constraints, oil spills, or a direct decline in oil supply, in turn, indicate a possible oil shortage due to the negative impact of these events on supply. 
       Likely shortages can also be inferred from news indicating a rise in demand, an increase in imports, or a drop in exports. 
       Generally, news that signals a scarcity of oil or a price increase should have a positive impact on the price. 
\item \emph{No price changes (score: $0$):} A neutral score has been assigned to the relatively small number of headlines that report no signs of supply, demand, imports, or exports. 
\end{itemize}

\subsubsection{Model fine-tuning}
The labeled headlines with the corresponding domain-specific sentiment scores yielded the S\&D-Dataset which contains approximately 14,000 negative, 500 neutral, and 15,000 positive headlines. 
We split the dataset into training (60\%), test (20\%), and validation (20\%) partitions (keeping the distribution across classes), and used the test dataset for fine-tuning FinBERT resulting in the CrudeBERT classifier (Figure~\ref{fig:process-crudebert}):

\begin{figure}[htb]
\begin{varwidth}{\linewidth}
    \resizebox{\columnwidth}{!}{\includegraphics[width=0.5\textwidth]{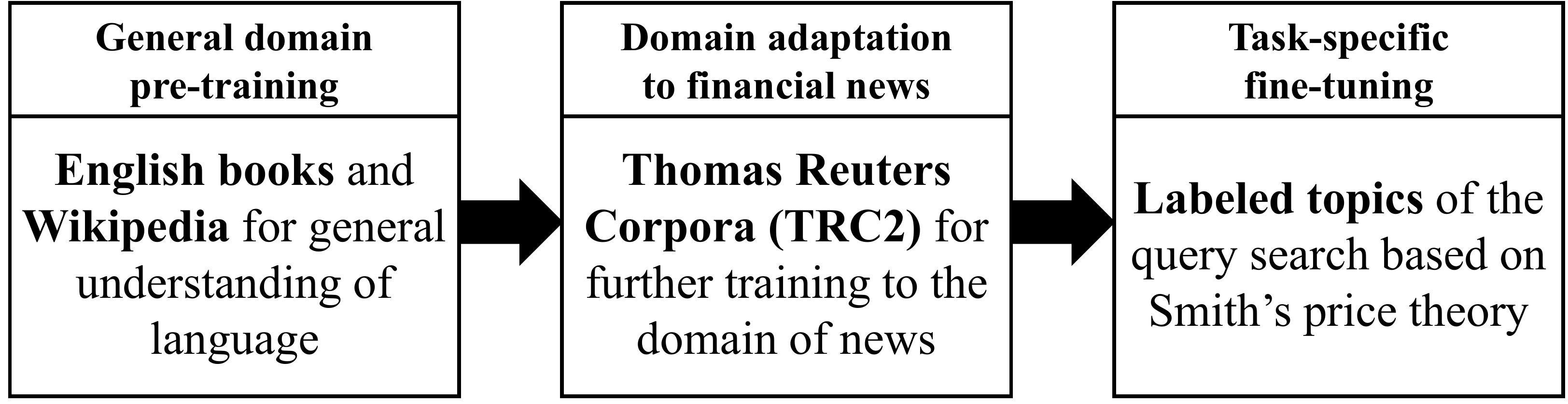}}
    \caption{\label{fig:process-crudebert}Process of Fine-tuning FinBERT to CrudeBERT.}
\end{varwidth}    
\end{figure}

Despite the relatively low number of neutral headlines, we included them in training to provide the neural network with examples of lower domain-specific sentiment scores that have not been assigned to one of the two extremes (i.e., $+1$ for positive and $-1$ for negative news).

\begin{figure}[H]
\centering
\subfigure[FinBERT]{
\includegraphics[width=.35\textwidth]{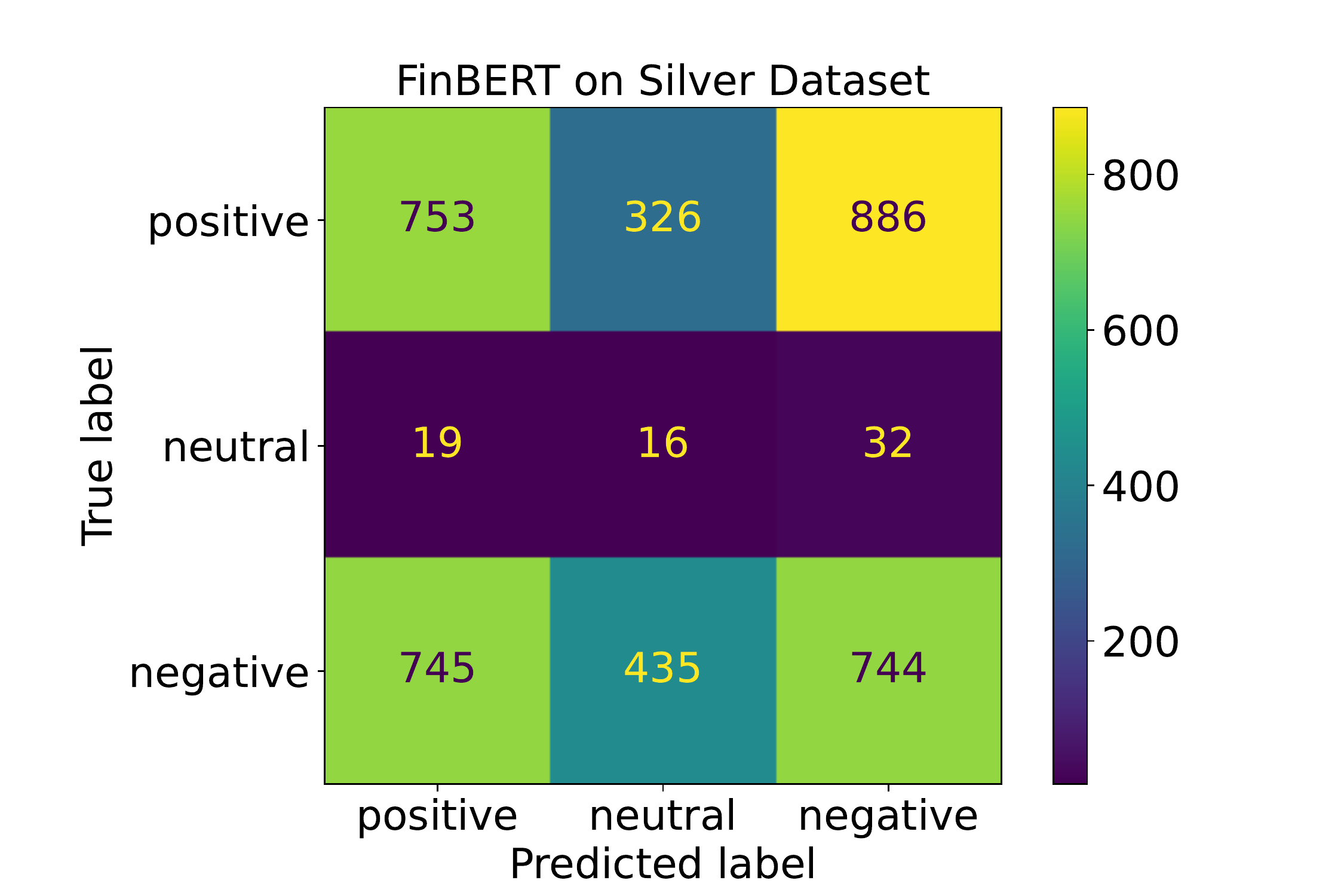}
}
\subfigure[CrudeBERT]{
\includegraphics[width=.35\textwidth]{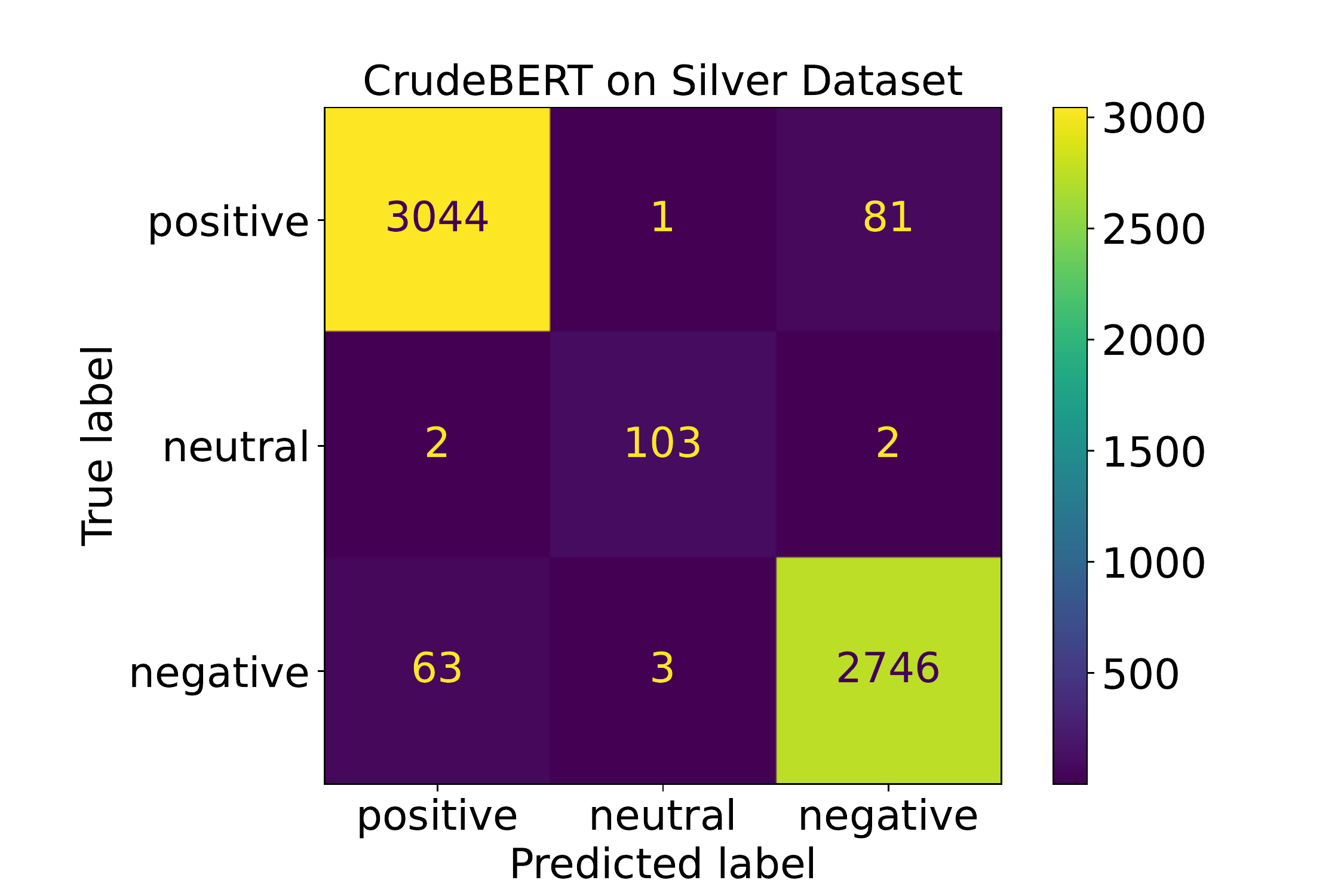}
}
\caption{\label{fig:Comparison SC on silver dataset}Confusion Matrices of the Two Transformer-based Financial Sentiment Classifiers on the Silver Dataset.}
\label{fig:whatever}
\end{figure}

A preliminary evaluation of the CrudeBERT classifier on the silver standard dataset yielded, despite the class imbalance, a macro F1 score of 0.97, a macro accuracy of 0.98, and a macro recall of 0.97 (Figure~\ref{fig:Comparison SC on silver dataset}). 
On the other hand, the same evaluation with the FinBERT classifier yielded a macro F1 score of 0.29, a macro accuracy of 0.59, and a macro recall of 0.32 on the silver test dataset (Figure~\ref{fig:Comparison SC on silver dataset}). 
Given the substantial amount of headlines used for fine-tuning and their relatively short length (on average 10.4 words per headline), these improvements are not surprising. 

\begin{figure}[htb]
\begin{varwidth}{\linewidth}
    \resizebox{\columnwidth}{!}{\includegraphics[width=0.5\textwidth]{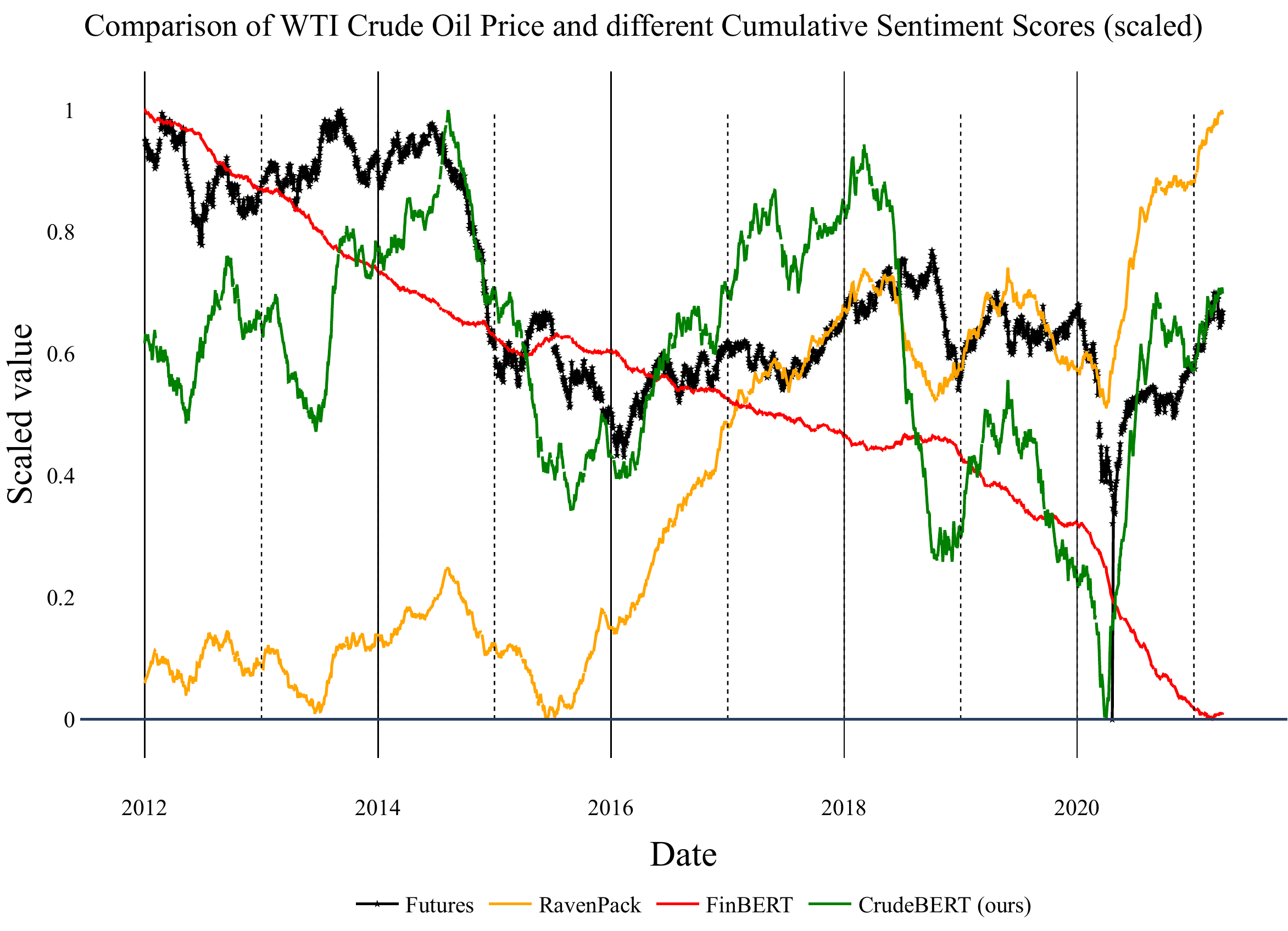}}
    \caption{Comparison of WTI Crude Oil Futures Prices and Different Cumulative Sentiment Scores (Scaled).}\label{fig:plot_cumulative_values}
\end{varwidth}    
\end{figure}

The qualitative comparison in Figure~\ref{fig:plot_cumulative_values} further supports our initial intuition that FinBERT's lack of knowledge of an event's impact on supply and demand seriously limits its suitability for prediction tasks. Consequently, it fails to track historical price movements compared to the fine-tuned CrudeBERT model and the commercial classifier of RavenPack.

\section{\uppercase{Evaluation}}
\label{sec:evaluation}
The following experiments leverage three different sentiment classifiers (FinBERT, CrudeBERT, and RavenPack ESS) to assess the potential of analyzing headlines for predicting the direction of the next day's ($Return_{t+1}$) change in crude oil futures prices, using a two-class higher/lower price classification schema.

The evaluation considers the period between 1 January 2012 and 1 April 2021 consisting of 3376 days' worth of data. We use precision, recall, and the F1 metric to assess the predictive potential of the evaluated classifiers.

Table~\ref{tab:Classification Report} summarizes the evaluation results. 
On average CrudeBERT outperforms FinBERT, RavenPack, and a random baseline for binary classification.
Applying FinBERT without any customizations to the prediction task seems to be contra-productive since it yields worse results than the random baseline. 
Fine-tuning FinBERT with the presented domain adaptation method considerably improves the method's performance. 
CrudeBERT's overall predictions also surpass the results from RavenPack's proprietary sentiment classifier, although these differences are less pronounced. 
CrudeBERT performs slightly worse for price-up predictions but considerably better at predicting pre-down movements.

\begin{table}[htb]
  \centering
  \caption{\label{tab:Classification Report}Classification Report of Different Sentiment Classifiers for Predicting Following Day WTI Oil Futures.}
  \resizebox{\columnwidth}{!}{%
  \begin{tabular}{lllllll} 
    \toprule
   {\thead{Metric}} & {\thead{Category}} & {\thead{FinBERT}} & {\thead{Random}} & {\thead{RavenPack}} & {\thead{CrudeBERT}}  \\
    \midrule
    Precision & Price down & 0.49 & 0.51   & 0.51  & 0.53\\
              & Price up   & 0.44 & 0.50   & 0.51  & 0.53\\
              & Macro      & 0.46 & 0.51   & 0.51  & \textbf{0.53} \\
              \midrule
    Recall    & Price down & 0.85 & 0.51   & 0.47  & 0.53\\
              & Price up   & 0.11 & 0.50   & 0.55  & 0.52\\
              & Macro      & 0.48  & 0.51   & 0.51  & \textbf{0.53}\\
              \midrule
    F1-Score  & Price down & 0.62 & 0.51   & 0.49  & 0.53\\
              & Price up   & 0.18 & 0.50   & 0.53  & 0.52\\
              & Macro      & 0.40  & 0.51   & 0.51  & \textbf{0.53}\\
    \bottomrule
  \end{tabular}
  }
\end{table}

\begin{figure*}[htb]
\centering
\subfigure[FinBERT]{
\includegraphics[width=.48\textwidth]{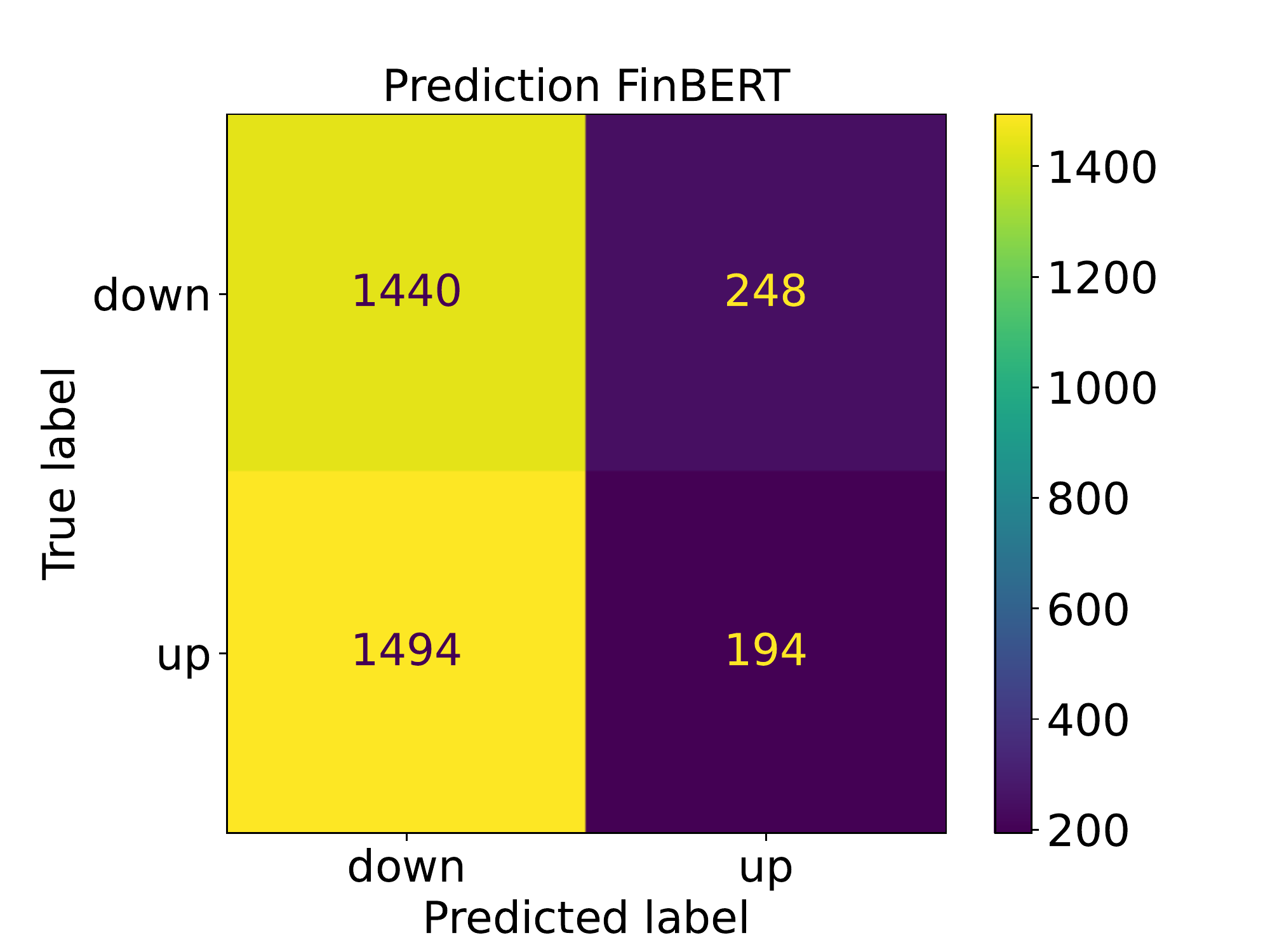}
}
\subfigure[Random]{
\includegraphics[width=.48\textwidth]{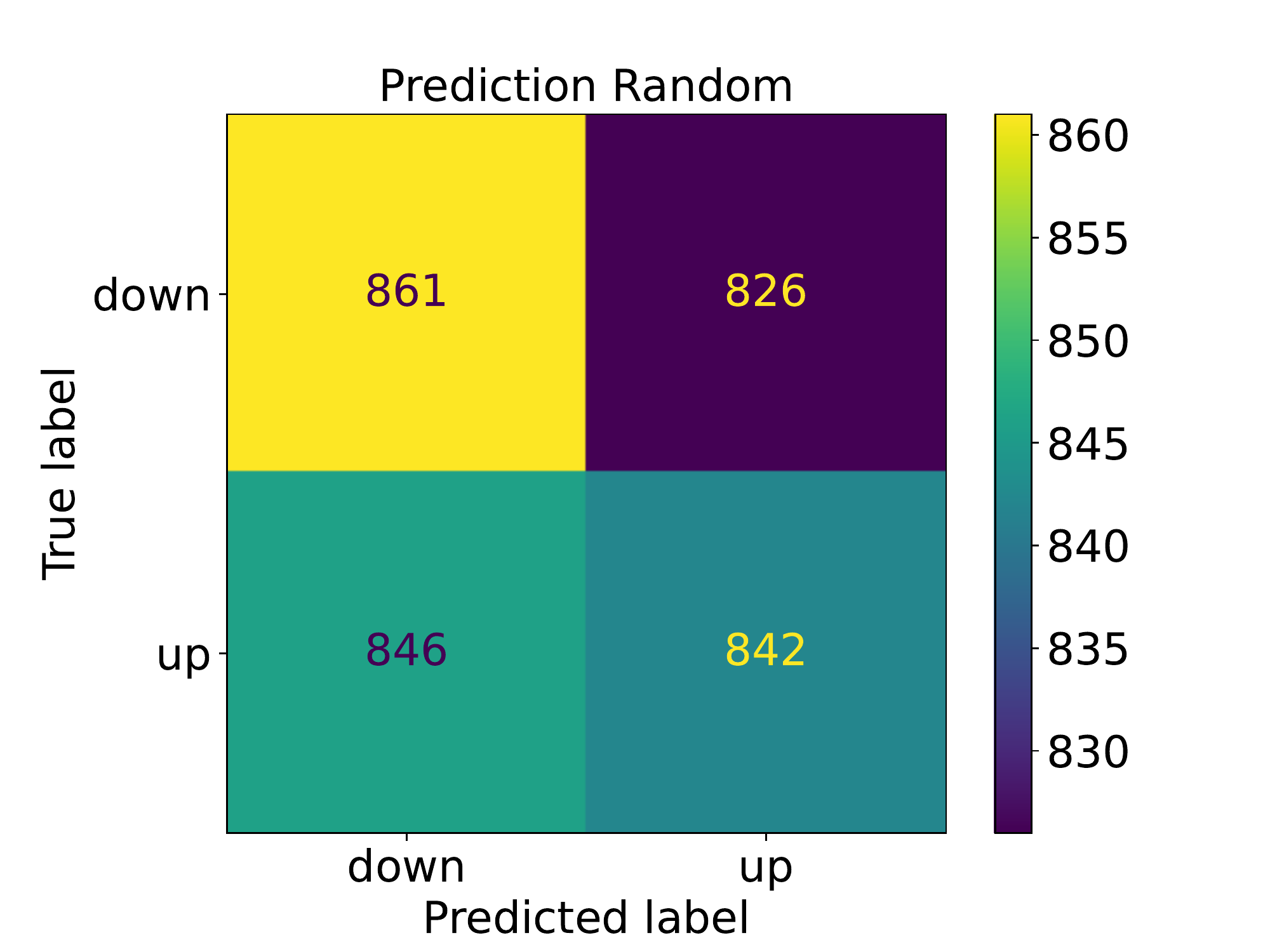}
}
\subfigure[RavenPack]{
\includegraphics[width=.48\textwidth]{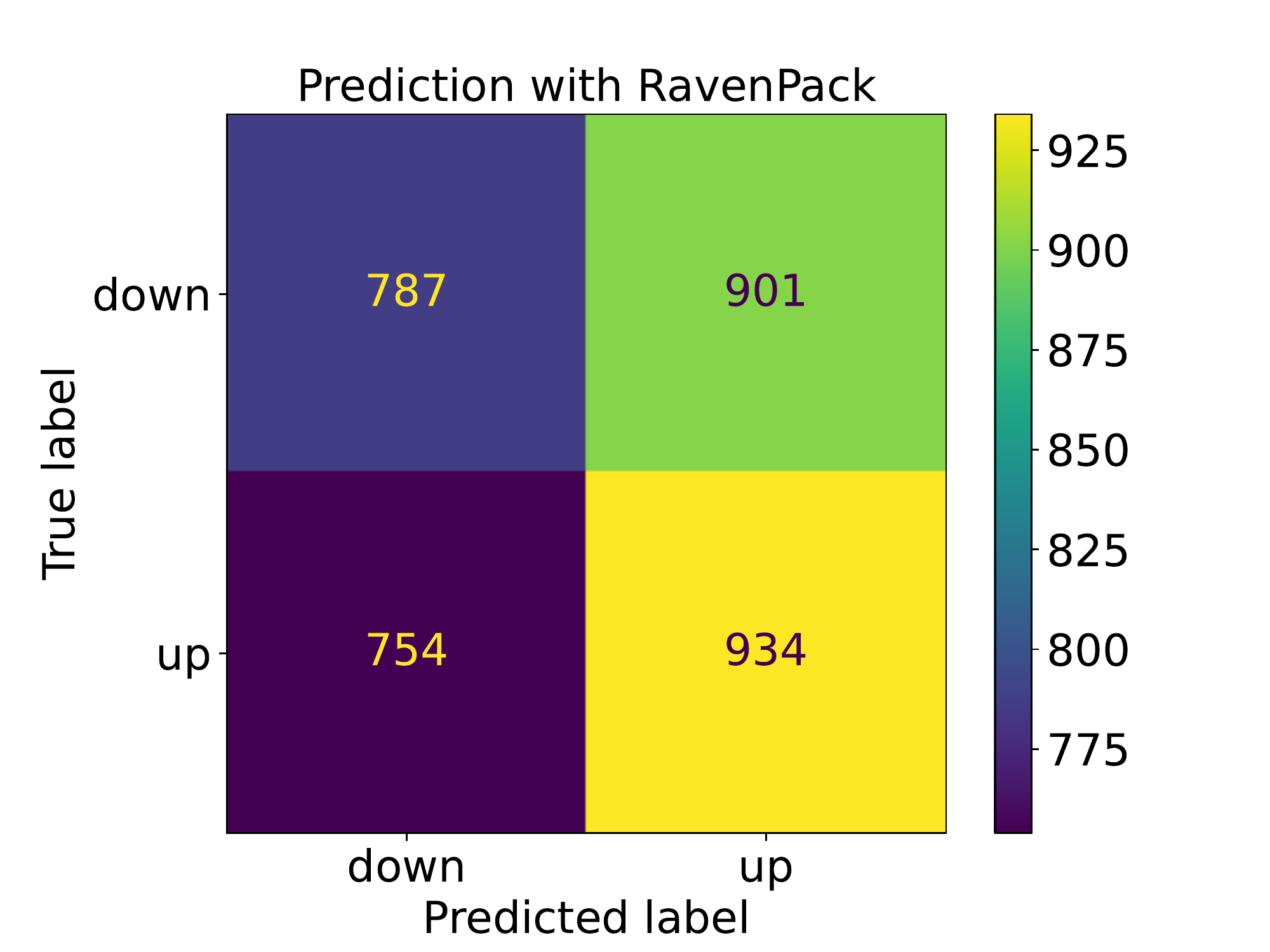}
}
\subfigure[CrudeBERT]{
\includegraphics[width=.48\textwidth]{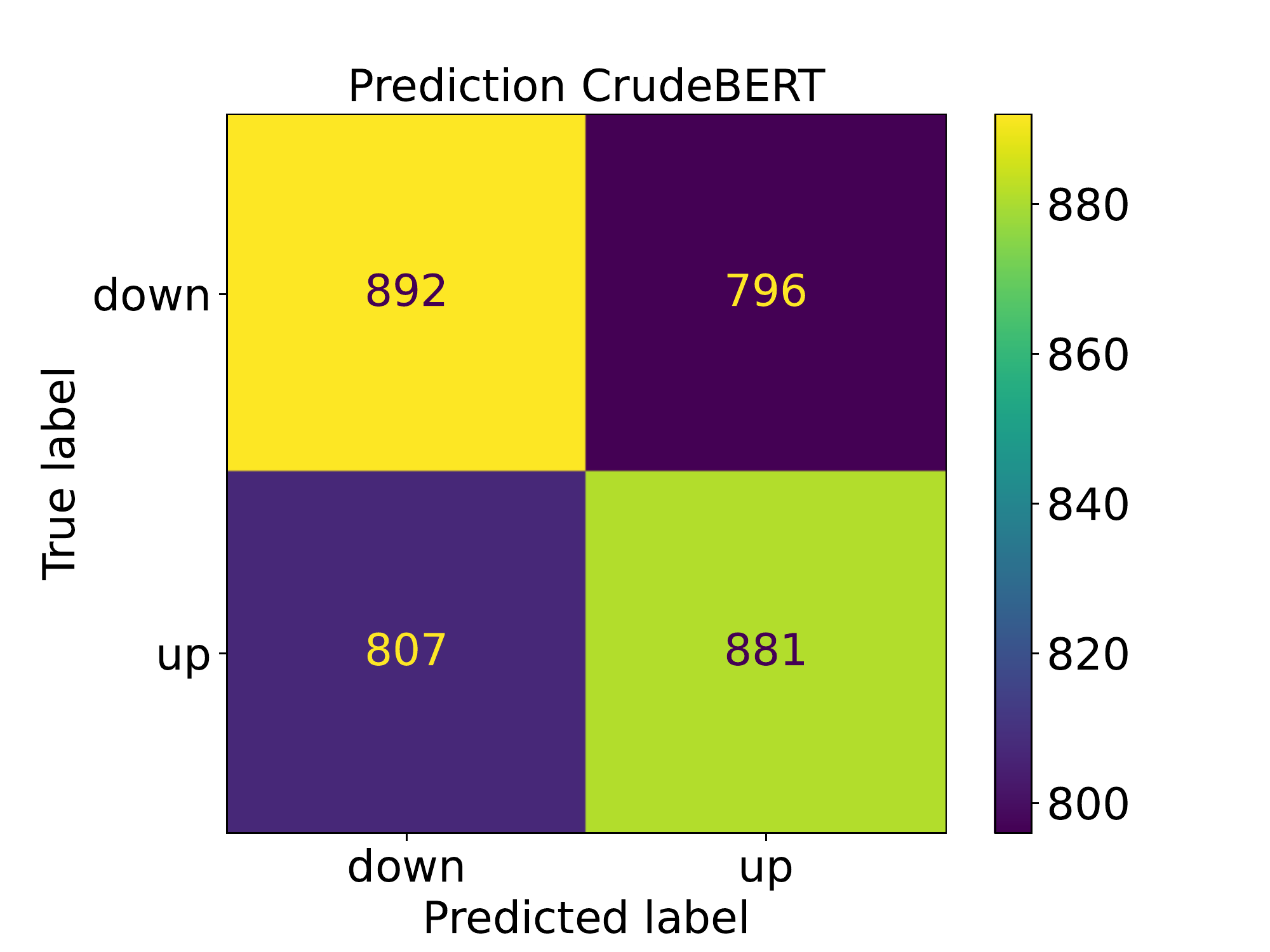}
}

\caption{\label{fig:Comparison Confusion Matrices} Confusion Matrices Comparing the Price Changes Predicted by RavenPack, FinBERT, and CrudeBERT with the Recorded Price Changes of the Following Day WTI Crude Oil Futures.}
\end{figure*}

Figure~\ref{fig:Comparison Confusion Matrices} presents a confusion matrix that compares the predicted label for each classifier with the following day's price changes of WTI crude oil futures ($Return_{t+1}$).

We, therefore, drew upon the SciPy\footnote{https://scipy.org} stats package to perform Pearson's chi-square test to determine whether the improvements provided by CrudeBERT are statistically significant. When compared to either FinBERT or the random baseline, both CrudeBERT and RavenPack yield significantly better results at the 0.05 significance level. 
The improvements from RavenPack to CrudeBERT (1721 versus 1773 correct predictions) are less substantial and have only been judged significant at the 0.10 significance level.



\section{\uppercase{Outlook and Conclusion}}
\label{sec:outlook}
Predicting market movements based on news headlines is still a very challenging task, as outlined in the experiments conducted in Section~\ref{sec:evaluation}. Even FinBERT, a state-of-the-art sentiment classifier that contains knowledge about the general financial domain, is unable to offer helpful insights into the future price fluctuations of commodities like crude oil when used without any domain adaptations.

The presented paper, therefore, introduces a method for fine-tuning FinBERT based on news headlines. Our approach selects frequently reoccurring topics that cover events illustrating fundamental market dynamics such as the interplay between supply and demand. A frequency analysis identifies these topics which are then used as keywords in search queries for collecting additional suitable headlines. 
Classifying the retrieved headlines based on Adam Smith's price theory allows the creation of a silver standard dataset, which serves as a practical and cost-effective alternative to human-curated training datasets. 
Applying this method to the domain of crude oil led to the creation of a silver standard that has then been used for fine-tuning FinBERT to create CrudeBERT, a domain-specific affective model that provides significantly better results than the original transformer model. In our experiments, which cover crude oil futures price movements over a nine-year period, CrudeBERT outperforms FinBERT and a random baseline on a significance level of 0.05. CrudeBERT even yields better results than RavenPack's proprietary sentiment analysis model which has been optimized in years of development, although the observed improvements are only significant on the 0.10 significance level.

Future research on evaluation methods and metrics will help to better understand the relationship between the model's predictions and future crude oil prices. The presented experiments only shed light upon its short-term prediction performance (i.e., $Return_{t+1}$ which covers the next business day). Thus, further research is required to investigate CrudeBERT's suitability for long-term strategies and in different economic environments (e.g., during times of economic boom or recession). 

It is also noteworthy that news headlines alone rather than the whole article seem to be sufficient for providing insights into the likely direction of price changes. Despite the presented improvements, CrudeBERT still has limitations and will be subject to further developments. We also intend to provide CrudeBERT with the ability to distinguish named entities (e.g. countries and oil companies) and numerical clues (e.g. \textit{increased by 10\,\%}  and \textit{increased by 1\,\%}) to obtain a more fine-grained sentiment score. This improved indicator should no longer be limited to providing information on the direction of price movements but also express their valence. Considering news volume seems to be another strategy for assessing an event's impact on the market.

Future research will also address the silver standard generation process. 
The current process, for instance, does not contain any additional logic for hand\-ling headlines with contradictory information on future supply and demand (e.g., ``Ivory Coast Jan Crude Oil Exports -1\,\% On Yr, Imports -3\,\%''). 
We, therefore, plan to develop strategies for identifying and processing such mixed-signal news headlines. 

Furthermore, we aim to assess the feasibility of extending the presented method to other commodities such as perishable (e.g., coffee beans), non-perishable (e.g., natural gas), precious (e.g., gold), and non-precious (e.g., iron ore) commodities, where pricing may be influenced by similar factors.

\section*{ACKNOWLEDGEMENT}
We would like to extend our gratitude to Prof Dr Hans Wernher van de Venn and the Institute of Mechatronic Systems at Zurich University of Applied Sciences for their generous support of this research.
In addition, we would like to thank Dr Adrian M.P. Bra\c soveanu for his valuable inputs on suitable evaluations for the CrudeBERT model. We would also like to thank Dr Sahand Haji Ali Ahmad for his assessment of the relevance of the news categories used in developing the silver standard dataset.



\bibliographystyle{apalike}

{\small

\end{document}